\documentstyle[12pt]{article}
\begin{document}       
\title{On the Fairlie's Moyal formulation of M(atrix)-theory}       
\author{\\M.Hssaini$^1$  ,  M.B.Sedra $^{1,2}$\\
\small{Abdus Salam International Centre For Theoretical Physics ICTP, Trieste, Italy}\\
\\{M.Bennai  \thanks{On leave of absence from: Faculty of Sciences, Ben M'Sik,
 Casa Morocco}}      and     B.Maroufi $^1$\\
 \small {$1$ Facult\'e des Sciences, D\'epartement de Physique, UFR-PHE}
 \small {Av. Ibn Battouta} \\
 \small {B.P. 1014, Rabat-Morocco}\\
 \small {$2$ Laboratoire de Physique Th\'eorique et Appliqu\'ee LPTA}
 \small {Facult\'e des Sciences}\\
 \small{D\'epartement de Physique, B.P. 133, Kenitra, Morocco}}

\maketitle
\hoffset=-1cm\textwidth=11,5cm                      
\vspace*{0.5cm}
     
\begin{abstract}

Starting from the Moyal formulation of M-theory in the large N-limit,
we propose to reexamine the associated membrane equations of motion in 10
dimensions formulated in terms of Poisson bracket. Among the results obtained,
we rewrite the coupled first order Nahm's equations into a simple form
leading in turn to their systematic relation with $SU(\infty)$ Yang Mills
equations of motion. The former are interpreted as the vanishing condition
of some conserved currents which we propose. We develop also an algebraic
analysis in which an ansatz is considered and find an explicit form for the
membrane solution of our problem. Typical solutions known in literature can also
emerge as special cases of the proposed solution.\\

\end{abstract}
\newpage
\section{Introduction}

Matrix model formulation of M-theory, introduced three years ago by Banks et al [1], 
and studied later by many authors[2-3], have shown to be an important area of research.
In this model, recall for instance that the fundamental degrees of freedom of the theory are 
the D0-branes whose interaction processus are described by space-time matrix coordinates[1].
Actually, we know that this matrix model is described by the maximally supersymmetric
U(N)Yang-Mills gauge theory, where N is the light-like momentum or the number of D0-branes interpreted in [2] as the number
of Green Schwartz strings (gas of N strings) in the light cone gauge. Its also known that these 
strings have differents lenghts depending on the winding number $m_k$ around the compact 
direction, a fact which can be described by the following expression\\
\begin{equation}
N=\sum{m_k}{N_k} .
\end{equation}
The large N limit is then shown to corresponds to long strings in the infinite momentum frame
(IFM). Several ways to investigate this large N limit have been followed. Among these issues one 
focus on the work of Fairlie [4] which purposes to clarify the meaning of this infinite limit
in 4 and 10 dimensions in terms of Moyal bracket formalism. As claimed by this author, this
Moyal bracket description is motivated by the fact that the matrix theory takes on an aspect
analogous to a Moyal formulation of quantum mechanics[4]. Note by the way, that the
Moyal bracket is defined via the star product known to be an essential ingredient towards setting
a non commutative geometry framework [5-6]. Recently this old mathematical idea have been a 
subject of a revived interest in the study of string and matrix theory [7-8-9-10].\\

Before going into the description of the main lines of the present work, we will try to
summarise the Fairlie's Moyal bracket formulation in M-theory as done in [4]. As signalled
above and in accordance with [4], the author examine the Moyal version of matrix theory in the
belief that this is the most appropriate for the discussion of the large N limit, and for 
the investigation of parallels with quantum mechanics. Indeed, it is consistently shown that 
the Moyal bracket description of infinite limit of matrix theory in 2 and 8 transverse 
dimensions leads to a class of solutions for the second order matrix theory equations 
obtained from the first order ones.\\

Using this correspondence between the second and the first
order equations, the author of [4] proceed then by solving the original theory by inverting
the problem which means finding a solution to first order equations which guarantees the 
integrability of the second order ones. This remarkable feature is due to the existence of the
 Bogomol'nyi bound to the Euclideanised version of the above equations in 4 and 10 dimensions.
The general construction of the obtained solutions is also given in terms of a representation
of the target space coordinates as non local spinor bilinears, which are generalisations of 
the standard wigner functions on phase space. \\

In the present work, we consider an alternative way to reexamine the matrix
theory equations of motion in 10 dimensions in the infinite limit, formulated
in terms of Poisson bracket instead of Moyal bracket. Among our results we
rewrite in a first step the Nahm's equations into a simple form. This consist
precisely in reducing the 14 coupled Nahm's equations (11,12) leading just to
five ones (13,14). These reduced equations are convenient in the sense that
their derivative with respect to $\sigma$ gives exactly the second order
equations of motion (16) which we derive starting from the $SU(\infty )$
pure Yang Mills lagrangian density. We show also that the second order
equations of motion exhibit a special invariance property as given in (21). \\

The next step consist to understand the meaning of the first order
Nahm's equations and their relation with the equations of motion. For
this, we give an interpretation of the former as the vanishing condition
of some conserved currents that we consider and interpret the above
correspondence as the conservation property of these currents. Finally,
we propose to solve the derived second order equations of motion. We
have accomplish this task by considering the Fourier modes
decomposition, using some algebraic manipulations and setting the
important ansatz (48) leading to build our membrane solution.\\

Next, we try to look for the possible connection of our solution with
other solutions known in literature. We start first from the important
remark that our membrane solution (71) exhibits an oscillation behaviour
which can be reduced to give rise to other solutions (see for example
[15]) once particular choices of our parameters are performed. \\

We organise this paper as follows. In section 2, we review the large N limit
in terms of Moyal bracket as done in [4] and present our alternative
approach in section 3 to derive the second order membrane equations of motion
in 10 dimensions. In section 4, we give a breif comment about Nahm's
equations and present in section 5 the analysis leading to construct a
membrane solution in 10 dimensions. Section 6 is devoted to our conclusion.

\section{Moyal bracket and large N limit }

Since matrix model formulation of M-theory was introduced by Banks et al [1], a lot of 
stimulated papers elaborating the issue was done [2-3]. Focusing for instance one of these
 papers [4], when the author describes the large N limit of matrix theory in 4 and 10 
dimensions using  the Moyal bracket formalism. The starting point is the matrix theory 
with an action of a two dimensional $N=8$ supersymmetric U(N) Yang-Mills theory
namely [2]\\
\begin{equation}
S={1\over{2\pi\alpha^\prime}}\int{Tr((D_{\mu}X)^{2}}+\theta^{T}\not{D}\theta + 
{g_{s}}^2{F_{\mu\nu}}^{2}-{1\over{{g_{s}}^2}}[X^{i},X^{j}]^2+{1\over
g_{s}}\theta^{T}\gamma_{i}[X^{i},\theta]d\sigma d\tau ,
\end{equation}
where $X^i$ and $(\theta_{L}^{\alpha},\theta_{R}^{\alpha})$ are the 8 scalar and 8
 fermionic fields respectively, which are $N\times{N}$ hermetian matrices transforming 
as the 8v vector and (8s, 8c) spinor representations of the SO(8) R-symmetry group of transversal 
rotations.\\

The coordinates $\sigma$ and $\tau$ with $0\leq{\sigma}\leq{2\pi}$
parametrise a cylinder (world sheet). \\
The passage to the large N limit consist in a first step in
considering the matrices $X^i$  as functions of two phase space variables $\alpha,\beta$
as well as  $\sigma,\tau$ such that $X^i\equiv{X^{i}(\alpha,\beta,\sigma,\tau)}$  .
The associated matrix elements may be regarded as the Fourier components of $X^i$ .
The second step consist in substituting commutators of the action (2) by the Moyal bracket [11] 
in the following way
\begin{equation}
\int{Tr[X^{i},X^{j}]}d\sigma\longrightarrow{1\over\lambda^2}\int
 {d\sigma d\alpha d\beta }\{X^{i},X^{j}\}_{MB}^2   ,
\end{equation}
where $\{X^{i},X^{j}\}_{MB}=\sin\{X^{i},X^{j}\}$  is the sine or Moyal bracket, 
with deformation parameter $\lambda$ defined as the imaginary part of the star product *. 
Recall that the star product of two functions $X^i$ and $X^j$ is defined as:
\begin{equation}
X^{i}\ast X^{j} =\lim\limits_{\stackrel{\alpha^\prime\to\alpha}{\beta^\prime\to\beta}}
\exp^{{i}\lambda(\partial_{\alpha}\partial_{\beta^\prime}
-\partial_{\alpha^\prime}\partial_{\beta})}X^{i}(\alpha,\beta,\sigma)X^{j}
(\alpha^\prime,\beta^\prime,\sigma)    .
\end{equation}                          
The Moyal bracket is then defined as the antisymmetric part of the star product such that
\begin{equation}
\{X^{i},X^{j}\}_{MB}=
\lim\limits_{\stackrel{\alpha^\prime\to\alpha}{\beta^\prime\to\beta}}
\sin\lambda(\partial_{\alpha}\partial_{\beta^\prime}-\partial_{\alpha^\prime}
\partial_{\beta})X^{i}(\alpha,\beta,\sigma)X^{j}
(\alpha^\prime,\beta^\prime,\sigma)  .
\end{equation}
As quoted in [4], the point of this construction is that in the limiting points 
 $\lambda\rightarrow{2\pi\over{N}}$, the Moyal bracket (5) reproduces the commutators 
of $N\times{N}$ matrices $X^i$(bosonic coordinates) through the association of the components
 $X^i_{mn}$  of $X^i$ with the Fourier modes of a function $X^{i}(\alpha,\beta,\sigma)$ 
periodic in $\alpha,\beta$ .
The remaining terms in the action (2) involving fermionic coordinates are replaced by
\begin{equation}
\begin{array}{lcr}
\{X^{\mu},\theta\}_{MB}&=&
\lim\limits_{\stackrel{\alpha^\prime\to\alpha}{\beta^\prime\to\beta}}
\sin\lambda(\partial_{\alpha}\partial_{\beta^\prime}-\partial_{\alpha^\prime}\partial_{\beta})
X^{i}(\alpha,\beta,\sigma)\theta
(\alpha^\prime,\beta^\prime,\sigma)\\
\{\theta^{T},\not{D}\theta\}_{MB}&=&
\lim\limits_{\stackrel{\alpha^\prime\to\alpha}{\beta^\prime\to\beta}}
\cos\lambda(\partial_{\alpha}\partial_{\beta^\prime}-\partial_{\alpha^\prime}\partial_{\beta})
\theta^{T}(\alpha,\beta,\sigma)\not{D}\theta
(\alpha^\prime,\beta^\prime,\sigma).
\end{array}
\end{equation}
Thus the action becomes
\begin{equation}
\begin{array}{lcr}
S_{MB}={1\over{2\pi\alpha^\prime}}\int((D_{\mu}X)^{2}+\cos\{\theta^{T},
\not{D}\theta\}+{g_{s}}^2Tr{F_{\mu\nu}}^{2}) d\alpha d\beta d\sigma d\tau
\\-\int(({1\over{\lambda {g_{s}}^2}}\sin\{X^{i},X^{j}\})^2-
{1\over{g_{s}}}\cos\{\theta^{T}\gamma_{i},{1\over\lambda}\sin\{X^{i},\theta\}\})d\alpha d\beta 
d\sigma d\tau  .
\end{array}
\end{equation}
 Moreover, once the following correspondence $\lambda\rightarrow{2\pi\over{N}}$ is performed, 
the final form of the action , in the large N-limit ($\lambda\rightarrow{0}$), is then
 expressed in terms of ordinary Poisson brackets as
\begin{equation}
\begin{array}{lcr}
S_{PB}&=&{1\over{2\pi\alpha^\prime}}\int((D_{\mu}X)^{2}+\theta^{T}
\not{D}\theta+{g_{s}}^2Tr{F_{\mu\nu}}^{2}) d\alpha d\beta d\sigma d\tau
\\&&{}-\int(({1\over{{g_{s}}^2}}\{X^{i},X^{j}\})^2-
{1\over{g_{s}}}\theta^{T}\gamma_{i}\{X^{i},\theta\})d\alpha d\beta d\sigma d\tau  .
\end{array}
\end{equation}
The obtained action defines a membrane SU($\infty$) pure Yang Mills theory.
Later on, we will consider the longitudinal as well as the timelike
coordinates in order to treat the system dynamically [4].

\section{Membranes in 10 dimensions: an alternative approach}

The aim of this section is to derive the equations of motion associated with
the SU($\infty$) Yang-Mills theory describing the membrane in 8 transverse
directions and show how these second order equations are connected to the first
order Nahm's equations. The latter's are shown to play a central role as they
provide a way to emphasise the similarity to the phase space formulation of
quantum mechanics [4].\\

A remarkable feature of the situation with 8 transverse directions is that,
due to the existence of a self-dual (antisymmetric) 4-tensor $T_{\mu\nu\rho\sigma}$ in 10 dimensions, the theory admits a class 
of solutions which we can obtain from a first order formulation. The Lagrangian density 
describing the 10-dimensional membrane have then the form of an SU($\infty$)
pure Yang-Mills theory (8).\\
The situation is further simplified to one of dependence upon only one of
the variables $\sigma$ (apart from the ${\alpha,\beta}$  dependence of the gauge potential
 $X^{i}(\alpha,\beta,\sigma)$ ). Using the gauge choice $X^0=constant$, the Lagrangian density
 is simply written as [4]
\begin{equation}
\begin{array}{lcr}
L={1\over{2}}(\partial_{\sigma}X^{\mu})^2+{1\over4}
\{X^{\mu},X^{\nu}\}^2  ,\mu=1,.....9  ,
\end{array}
\end{equation}
where $\partial_{\sigma}= \partial /\partial \sigma$ and where the dependence
upon $\tau$ is ignored. The symbol$\{X^{\mu},X^{\nu}\}$  deals with the
Poisson bracket given by
\begin{equation}
\begin{array}{lcr}
\{X^{\mu},X^{\nu}\}=\partial_{\alpha}X^{\mu}\partial_{\beta}X^{\nu}-
\partial_{\beta}X^{\mu}\partial_{\alpha}X^{\nu}   .
\end{array}
\end{equation}
Note by the way that (9) describes a 10-dimensional membrane with $X^0=constant$,
living inside 8 transverse directions. This situation is motivated by the fact that 
the matrix theory admits a class of solutions obtainable from a first order formulation 
(the Nahm equations). This is a set of 9 first order differential equations with 6 constraint 
equations given by
\begin{equation}
\begin{array}{lcr}
\partial_{\sigma}X^1+\{X^2,X^9\}=0\\
\partial_{\sigma}X^2+\{X^9,X^1\}=0\\
\partial_{\sigma}X^3+\{X^4,X^9\}=0\\
\partial_{\sigma}X^4+\{X^9,X^3\}=0\\
\partial_{\sigma}X^5+\{X^6,X^9\}=0\\
\partial_{\sigma}X^6+\{X^9,X^5\}=0\\
\partial_{\sigma}X^7+\{X^8,X^9\}=0\\
\partial_{\sigma}X^8+\{X^9,X^7\}=0\\
\partial_{\sigma}X^9+\{X^1,X^2\}+\{X^3,X^4\}+\{X^5,X^6\}+\{X^7,X^8\}=0   ,
\end{array}
\end{equation}
with
\begin{equation}
\begin{array}{lcr}
\{X^1,X^3\}+\{X^4,X^2\}+\{X^5,X^7\}+\{X^8,X^6\}=0\\
\{X^1,X^4\}+\{X^2,X^3\}+\{X^8,X^5\}+\{X^7,X^6\}=0\\
\{X^1,X^5\}+\{X^4,X^8\}+\{X^7,X^3\}+\{X^6,X^2\}=0\\
\{X^1,X^6\}+\{X^2,X^5\}+\{X^3,X^8\}+\{X^4,X^7\}=0\\
\{X^1,X^7\}+\{X^3,X^5\}+\{X^8,X^2\}+\{X^6,X^4\}=0\\
\{X^1,X^8\}+\{X^5,X^4\}+\{X^2,X^7\}+\{X^6,X^3\}=0 .
\end{array}
\end{equation}
Focusing for the moment to reduce much more these equations, we have used some algebraic
manipulations and showed that (11-12) can be simply written in the following way
\begin{equation}
\begin{array}{lcr}
\partial_{\sigma}X^{\mu}+(-)^{\mu+1}\{X^{\mu+(-)^{\mu+1}},X^9\}=0,  \mu=1,...,8 &(a)\\
\partial_{\sigma}X^9+\sum_{\mu=1}^8{1\over2}(-)^{\mu+1}\{X^{\mu},X^{\mu+(-)^{\mu+1}}\}=0, 
  &(b)
\end{array}
\end{equation}

 and
\begin{equation}
\begin{array}{lcr}
\{X^1,X^{j}\}+(-)^{\j+1}\{X^{\j+(-)^{\j+1}},X^2\}=\{X^7,X^{\j+2(-)^{\j+1}}\}+(-)^{\j+1}\{X^{\j+3(-)^{\j+1}},X^8\},j=3,6\\
\{X^1,X^{j}\}+(-)^{\j+1}\{X^{\j+(-)^{\j+1}},X^2\}=\{X^5,X^{\j-4(-)^{\j+1}}\}+(-)^{\j+1}\{X^{\j-3(-)^{\j+1}},X^6\},j=4,7\\
\{X^1,X^{j}\}+(-)^{\j+1}\{X^{\j+(-)^{\j+1}},X^2\}=\{X^3,X^{\j+2(-)^{\j+1}}\}+(-)^{\j+1}\{X^{\j+3(-)^{\j+1}},X^4\},j=5,8
\end{array}
\end{equation}
Indeed, setting for example $\mu=1,2$ , we recover respectively from (13-a) the first two 
equations of (11) namely
\begin{equation}
\begin{array}{lcr}
\partial_{\sigma}X^1+\{X^2,X^9\}=0\\
\partial_{\sigma}X^2+\{X^9,X^1\}=0.
\end{array}
\end{equation}
Next note the important remark that there exist an intriguing correspondence between
the first order Nahm's equations (13) and the second order derived equations of motion.
Indeed consider the lagrangian (9); and applying the Euler Lagrange equations we find
the equations of motion for the membrane, namely:
\begin{equation}
\begin{array}{lcr}
\partial_{\sigma}^2X^{\mu}+\sum_{\nu=1}^8\{X^{\nu},\{X^{\nu},X^{\mu}\}\}=0,
\end{array}
\end{equation}
with $\mu=1,...,9$ and $X_0 =constant$. The above correspondence is then established by
derivating the Nahm's equations with respect to the coordinate $\sigma$ and using the
 constraints (14). As an example consider (13.b) such that
 \begin{equation}
\begin{array}{lcr}
\partial_{\sigma}^2X^9+\sum\limits_{\nu=1}^8{1\over2}(-)^{\nu+1}(\{\partial_
{\sigma}X^{\nu},X^{\nu+(-)^{\nu+1}}\}+\{X^{\nu},\partial_{\sigma}X^{\nu+(-)^{\nu+1}}\})=0,
\end{array}
\end{equation}
which becomes upon using (13.a)
\begin{equation}
\begin{array}{lcr}
\partial_{\sigma}^2X^9+ \sum\limits_{\nu=1}^8{1\over 2}(\{X^{\nu+
(-)^{\nu+1}},\{X^{\nu+(-)^{\nu+1}},X^9\}\}+\{X^{\nu},\{X^{\nu},X^9\}\})=0,
\end{array}
\end{equation}
and reproducing then exactly the equations of motion (16) for $\mu=9$ with the following algebraic 
property
\begin{equation}
\begin{array}{lcr}
\sum\limits_{\nu=1}^8\{X^{\nu+(-)^{\nu+1}},\{X^{\nu+(-)^{\nu+1}},X^9\}\}=
\sum\limits_{\nu=1}^8\{X^{\nu},\{X^{\nu},X^9\}\}.
\end{array}
\end{equation}
Note by the way that by virtue of this equality, the equations of motion (16)
can be equivalently written as:
\begin{equation}
\begin{array}{lcr}
\partial_{\sigma}^2X^{\mu}+\sum\limits_{\nu=1}^8\{X^{\nu+
(-)^{\nu+1}},\{X^{\nu+(-)^{\nu+1}},X^{\mu}\}\}=0,
\end{array}
\end{equation}
which shows an invariance property with respect to the following transformation
\begin{equation}
\nu\rightarrow \nu+(-)^{\nu+1}\equiv \nu+\exp^{i\pi{(\nu+1})} ,
\end{equation}
affecting the repeated index $\nu$ .\\

Consequently we note that the equation of motion concerning the longitudinal
coordinates $X^9$ is simply obtained when derivating with respect to $\sigma$.
The remaining second order equations of motion, for transverse directions
$\mu=1,...,8$, arise thanks to the existence of constraint equations.

\section {About the Nahm's equations}

Having shown explicitly how the second order equations of motion are derived
from the Nahm's first order ones, we will try now to search for the meaning
of the relation between these two kind of equations. For this task, we will
not ignore for instance about the coordinate $\tau$ and assume that our fields
$X^{\mu}$ depend on the full set of coordinates $\alpha, \beta, \sigma$ and
$\tau$ defining a space containing the world volume of the membrane.\\

This space can be structured as follows. Let $\sigma$ and $\tau$ define a
complex two-dimensional world-sheet $\Sigma$ with local coordinates $Z=\sigma+
i\tau$ and ${\bar Z}=\sigma-i\tau$ .\\ We use $\partial$ and
$ \bar \partial$ to denote $ \partial /{\partial z}$ and $ \partial /{\partial {\bar z}}$ respectively. The coordinates
$\alpha, \beta$ parametrise a phase space which we denote by $P (\alpha, \beta)$ such that
the fields $X^{\mu}$ are shown to live inside the space $\Sigma \otimes P(\alpha, \beta)$.\\
The lagrangian describing the SU($\infty$) Yang-Mills theory of the membrane
can be written formally as\\
\begin{equation}
\begin{array}{lcr}
L={1\over{2}}(\partial X^{\mu})({\bar \partial} X^{\mu})+{1\over4}\{X^{\mu},X^{\nu}\}^2  ,\mu=1,..9  ,
\end{array}
\end{equation}
in such a way that for $\tau =0$ , one recover directly the standard theory
 (9) for which $ \partial = \bar{\partial}= \partial_ {\sigma} $ .\\
 The equation of motion associated to the lagrangian (22) is given by \\
\begin{equation}
\begin{array}{lcr}
\partial \bar {\partial} X^{\mu}+\sum_{\nu=1}^8\{X^{\nu},\{X^{\nu},X^{\mu}\}\}=0 .
\end{array}
\end{equation}
This equation looks like a standard 2d conformal field theory equation of
motion, for which we are usually interested in deriving the conserved currents,
discussing the underlying conformal symmetries and integrability. A part from
being interesting for the above reasons, our equations contains further
informations concerning the structure of the membrane in 8 transverse
directions and the "matrix" behaviour of the fields $X^{\mu}$.\\

Using our knowledge on conformal fields theories and integrable systems for
which 2d conserved currents are defined such that their conserved law
reproduces in some sense the equation of motion, we are for instance
interested in cheking what happens for our equation (23). In fact, we remark
from our previous analysis that its possible to associate conserved currents
to (23). Denoting these objects by $J^\mu \equiv ( J^i , J^9), i=1,...,8 $; we can write
\begin{equation}
\begin{array}{lcr}
J^i = \partial X^i + (-)^{i+1} \{X^{i+(-)^{i+1}} , X^9 \},  i = 1,...,8  &(a)\\
J^9 = \partial X^9 + \sum_{\nu=1}^8 {1\over2}(-)^{\nu+1}\{X^{\nu},X^{\nu+(-)^{\nu+1}}\}, &(b)
\end{array}
\end{equation}
and their conservation properties are \\
\begin{equation}
\begin{array}{lcr}
\bar {\partial} J^i =\partial \bar{ \partial} X^i + \sum_{\nu=1}^8\{X^{\nu},\{X^{\nu},X^i \}\}=0 , i = 1,...,8
 \end{array}
\end{equation}

and \\
\begin{equation}
\begin{array}{lcr}
\bar {\partial} J^9=\partial \bar {\partial} X^9 + \sum_{\nu=1}^8\{X^{\nu},\{X^{\nu},X^9 \}\}=0,
\end{array}
\end{equation}
by virtue of (23).  \\
Now, the point is that the conserved currents $J^{\mu}, \mu=1,...9$ given by
(24) are nothing but the objects defining the first order Nahm's equations
 (13) which we can rewrite as follows\\
\begin{equation}
\begin{array}{lcr}
J^{\mu}=0 ,  \mu=1,..8\\
J^9=0  .
\end{array}
\end{equation}
This property can be traced to the fact that when the coordinate $\tau$ is
ignored, which means setting simply $\tau=0$, the currents $J^{\mu}
(\mu=1,...9$) become then vanishing objects and give rise then to the first
order Nahm's equations (13). We can then interpret these kind of equations,
as the vanishing property of the conserved currents of the theory (22) and
interpret the conservation law equations (25, 26) as the property giving the
correspondence between the first and the second order equations as discussed
previousely.

\section{Solving the equations of motion}

In this section we will return back to our discussion of section 3 in which
we ignore the parameter $\tau$ and consider the following equations of
motion(16)
\begin{equation}
\begin{array}{lcr}
\partial_{\sigma}^2X^{\mu}+\sum\limits_{\nu=1}^8\{X^{\nu},\{X^{\nu},X^{\mu}\}\}=0.
\end{array}
\end{equation}
Now, having shown how these equations describe the membrane in 8 transverse
directions, we are now in position to look for the explicit solution of the
model. To start, we assume that the coordinates $X^{\mu}(\alpha,\beta,\sigma)$
can be written in terms of the Fourier modes as follows
\begin{equation}
\begin{array}{lcr}
X^{\mu}(\alpha,\beta,\sigma)=\sum\limits_{mn\in{Z}}X^{\mu}_{mn}(\sigma)L_{mn}(\alpha,\beta) ,
\end{array}
\end{equation}
which suppose the periodicity of $X^{\mu}(\alpha,\beta,\sigma)$ in $\alpha,\beta$
 \begin{equation}
\begin{array}{lcr}
X^{\mu}(\alpha+2\pi,\beta+2\pi,\sigma)=X^{\mu}(\alpha,\beta,\sigma)  ,
\end{array}
\end{equation}
with
\begin{equation}
\begin{array}{lcr}
L_{mn}(\alpha,\beta)=\exp{i(m\alpha+n\beta)} .
\end{array}
\end{equation}
Furthermore, the modes $X^{\mu}_{mn}$ as we will show bellow, can be considered as operator 
entries of the coordinates $X^{\mu}$ satisfying
\begin{equation}
\begin{array}{lcr}
\{X^{\mu},X^{\nu}\}=\sum\limits_{\stackrel{m_1,n_1}{m_2,n_2}}X^\mu_{m_1n_1}(\sigma)X^\nu_{m_2n_2}(\sigma)\{L_{m_1n_1},L_{m_2n_2}\},
\end{array}
\end{equation}
where
\begin{equation}
\begin{array}{lcr}
\{L_{m_1n_1},L_{m_2n_2}\}=(m_2n_1-m_1n_2)L_{m_1+m_2,
n_1+n_2} ,
\end{array}
\end{equation}
which reproduces in some sense the Poisson Bracket algebra in the large N limit
of the area preserving diffeomorphism on the torus [12]. Note by the way that
(33) share a striking resemblance with the Antoniadis et al Lie algebra [13].
This is an infinite dimensional algebra which was generalised to include the
Virasoro algebra, the Frappat et al symmetries [12] as well as their W and
central charges extensions [14].\\
An important question about (33) is to look for the meaning of the corresponding 
generalisations in our context. Moreover, the appearence of this algebra
indicates also how the SU(N) symmetry (Moyal) of the supersymmetric Yang-Mills
matrix theory becomes the area preserving diffeomorphism group namely
$SU(\infty)$ describing the membrane. \\

Now, before solving explicitly our problem, we will present some algebraic
properties related to (28). To start, consider the antisymmetry property
of the Poisson bracket (32) namely
\begin{equation}
\begin{array}{lcr}
\{X^{\mu},X^{\nu}\}=-\{X^{\nu},X^{\mu}\} ,
\end{array}
\end{equation}
from which we can write
\begin{equation}
\begin{array}{lcr}
\{X^{\mu},X^{\nu}\}& =&\sum\limits_{\stackrel{m_1,n_1}{m_2,n_2}}X^\mu_{m_1n_1}
(\sigma)X^\nu_{m_2n_2}(\sigma)\{L_{m_1n_1},L_{m_2n_2}\}\\

& =&-\sum\limits_{\stackrel{m_1,n_1}{m_2,n_2}}X^\nu_{m_1n_1}(\sigma)X^\mu_{m_2n_2}
(\sigma)\{L_{m_1n_1},L_{m_2n_2}\}.
\end{array}
\end{equation}
On the other hand
\begin{equation}
\begin{array}{lcr}
\{X^{\mu},X^{\nu}\}&=&\sum\limits_{\stackrel{m_1,n_1}{m_2,n_2}}X^\mu_{m_1n_1}
(\sigma)X^\nu_{m_2n_2}(\sigma)\{L_{m_1n_1},L_{m_2n_2}\}\\                   & =&-\sum\limits_{\stackrel{m_1,n_1}{m_2,n_2}}X^\mu_{m_1n_1}
(\sigma)X^\nu_{m_2n_2}(\sigma)\{L_{m_2n_2},L_{m_1n_1}\}\\                 &=&-\sum\limits_{\stackrel{m_1,n_1}{m_2,n_2}}X^\mu_{m_2n_2}
(\sigma)X^\nu_{m_1n_1}(\sigma)\{L_{m_1n_1},L_{m_2n_2}\} .
 \end{array}
\end{equation}
The above formulas, lead then to set the following results
\begin{equation}
\begin{array}{lcr}
X^{\mu}_{m_1n_1}(\sigma)X^{\nu}_{m_2n_2}(\sigma)=-X^{\nu}_{m_1n_1}(\sigma)X^{\mu}_{m_2n_2}(\sigma) ,
\end{array}                                                                                       
\end{equation}
\begin{equation}
\begin{array}{lcr}
X^{\mu}_{m_1n_1}(\sigma)X^{\nu}_{m_2n_2}(\sigma)=-X^{\mu}_{m_2n_2}(\sigma)X^{\nu}_{m_1n_1}(\sigma) ,
\end{array}
\end{equation}    
showing the antisymmetry feature of the components $X^{\nu}_{mn}$ with respect to both the 
indices $\mu,\nu$ and the combined index $(m_in_i)$, i=1,2 once the following mapping are
performed
\begin{equation}
\begin{array}{lcr}
(m_1n_1)&\longleftrightarrow&(m_2n_2)\\  \mu&\longleftrightarrow&\nu .
\end{array}
\end{equation}
Indeed, combining (37) and (38) we obtain
\begin{equation}
\begin{array}{lcr}
X^{\mu}_{m_1n_1}(\sigma)X^{\nu}_{m_2n_2}(\sigma)=X^{\nu}_{m_2n_2}(\sigma)X^{\mu}_{m_1n_1}(\sigma).
\end{array}
\end{equation}    
This result  shows explicitly the bosonic behaviour of the coordinates 
 $X^{\nu}_{mn}$ , a natural property which is suspected already at the level of the Fourier
 modes decomposition (29). Indeed, the latter's suppose from the begining that the coordinates
$X^{\nu}$ are "functions" of two phase space variables $\alpha,\beta$ as well as $\sigma$.\\

In the spirit to understand much more the equations of motion and their underlying
symmetries, we remark that a realisation of the component $X^{\nu}_{mn}$ is provided by the following expression  
\begin{equation}
\begin{array}{lcr}
X^{\mu}_{mn}(\sigma)={\gamma^{\mu}}X_{mn}(\sigma) ,
\end{array}
\end{equation}    
where ${\gamma^{\mu}}$, $\mu=1,...,9$ are the gamma matrices satisfying the Clifford algebra   
\begin{equation}
\begin{array}{lcr}
{\gamma^{\mu}}{\gamma^{\nu}}=-{\gamma^{\nu}}{\gamma^{\mu}}\\
{\gamma^{\mu}}^2=-1 .
\end{array}
\end{equation}    
Injecting for example (41) into (40), one find by virtue of (42) a non commutative property
 of the component $X_{mn}$ namely:   
\begin{equation}
\begin{array}{lcr}
X_{m_1n_1}(\sigma)X_{m_2n_2}(\sigma)=-X_{m_2n_2}(\sigma)X_{m_1n_1}(\sigma).
\end{array}
\end{equation}
In summary we can write (29) as follows    
\begin{equation}
\begin{array}{lcr}
X^{\mu}(\alpha,\beta,\sigma)=\sum\limits_{m,n\epsilon{Z}}{\gamma^{\mu}}X_{mn}
(\sigma)L_{mn}(\alpha,\beta) ,
\end{array}
\end{equation}                           
which gives after some computations
\begin{equation}
\begin{array}{lcr}
\sum\limits_{\nu=1}^{8}\{\{X^{\mu},X^{\nu}\},X^{\nu}\}=\sum\limits_{\nu=1}^{8}\sum\limits_{\stackrel{m_i,n_i}{i=1,2,3}}X_{m_1n_1}^{\mu}(\sigma)X_{m_2n_2}^{\nu}(\sigma)
X_{m_3n_3}^{\nu}(\sigma)\{\{L_{m_1n_1},L_{m_2n_2}\},L_{m_3n_3}\},
\end{array}
\end{equation}                           
with  
\begin{equation}
\begin{array}{lcr}
\{\{L_{m_1n_1},L_{m_2n_2}\},L_{m_3n_3}\}=(m_2n_1-m_1n_2)[m_3(n_1+n_2)
-n_3(m_1+m_2)]L_{\sum\limits_{i=1,2,3}m_i,\sum\limits_{i=1,2,3}n_i},
\end{array}
\end{equation}
and
\begin{equation}
\begin{array}{lcr}
\sum\limits_{\nu=1}^{8}X_{m_1n_1}^{\mu}(\sigma)X_{m_2n_2}^{\nu}(\sigma)X_{m_3n_3}^{\nu}(\sigma)=-8{\gamma}^{\mu}X_{m_1n_1}(\sigma)X_{m_2n_2}(\sigma)X_{m_3n_3}
(\sigma) .
\end{array}
\end{equation}                                               
Recall that our aim is to solve the membrane equations of motion which
consist to give an explicit solution for the coordinates $X_{mn}$ satisfying
the non commutative rule (43). In order to solve the above problem we propose
the following ansatz.
\begin{equation}
\begin{array}{lcr}
X_{m_1n_1}(\sigma)X_{m_2n_2}(\sigma)X_{m_3n_3}(\sigma)= {\epsilon_
{123}} F_{sym}(\{m_i\}, \{n_i\}) X_{m_1+m_2+m_3,n_1+n_2+n_3}(\sigma),
\end{array}
\end{equation}
where
\begin{equation}
\begin{array}{lcr}
{\epsilon_{123}}={\epsilon_{(m_1n_1)(m_2n_2)(m_3n_3)}} ,
\end{array}
\end{equation}                 
stands for the Levi-Cevita tensor given by
\begin{equation}
{\epsilon_{123}}={\large\{}
\begin{tabular}{ll}
-1 & for odd permutation \\
+1 & for even permutation . \\
\end{tabular}
\end{equation}
and where $ F_{sym}(\{m_i\}, \{n_i\}) $ is for the moment an arbitrary
function required to be symmetric with respect to the following permutation
of integer values
\begin{equation}
\begin{array}{lcr}
(m_in_i) \leftrightarrow (m_jn_j) , i, j =1, 2, 3   .
\end{array}
\end{equation}
Later on, we will denote simply this symmetric function as $F_{(123)}$.\\

The principal idea in setting the ansatz (48) is based on the fact that we
need to write the bi-Poisson bracket $\{\{X^{\mu},X^{\nu}\},X^{\nu}\}$ as a
simple function of $X_{mn}$ a fact which means that we should linearise the
cubic matrix product $X_{m_1n_1}X_{m_2n_2}X_{m_3n_3}$ to give rise to (48).
The apparition of the fully antisymmetric tensor ${\epsilon_{123}}$ in this
ansatz is justified by the non commutative behaviour of the components
$X_{mn}$ as given by (43). On the other hand, the function $F_{sym}$ introduced
in this ansatz and which we will discuss later, is chosen to be symmetric
in agreement with the above non commutativity property.
                                                
Actually with this ansatz; one can easily identify both the left and hand
sides terms of (28). In order to do this, remark first that we have
\begin{equation}
\begin{array}{lcr}
\sum \limits_{\nu=1}^{8}\{\{X^{\mu},X^{\nu}\},X^{\nu}\} = \\
\sum \limits_{\stackrel{m_i,n_i}{i=1,2,3}}
(- \gamma^{\mu}) 8 {\epsilon_{123}} F_{(123)}
{\omega}(\{m_i\},\{n_i\}) X_{\sum \limits_{i=1,2,3} m_i,
\sum \limits_{i=1,2,3} n_i}(\sigma)
L_{\sum\limits_{i=1,2,3}m_i,\sum\limits_{i=1,2,3}n_i} ,
\end{array}
\end{equation}
with
\begin{equation}
\begin{array}{lcr}
{\omega}(\{m_i\},\{n_i\})=(m_2n_1-m_1n_2)[m_3\sum\limits_{i=1}^3n_i-n_3
\sum\limits_{i=1}^3m_i] .
\end{array}
\end{equation}
which leads then to write the membrane equations of motion as follows
\begin{equation}
\begin{array}{lcr}
{\sum \limits_{p,q}} L_{p,q} \partial_{\sigma}^2 X_{p,q}=
\sum\limits_{\stackrel{m_i,n_i}{i=1,2,3}}
8{\epsilon_{123}} F_{(123)} {\omega}(\{m_i\},\{n_i\})X_{{\sum{m_i}},
{\sum {n_i}}}(\sigma) L_{\sum{ m_i},\sum{n_i}}  .
\end{array}
\end{equation}
Now, in order to do identification in (54) in a consistent way, one should
sum over the same indices $p, q$ in both the sides of this equation.
This is possible, since we can set $ p=\sum m_i $ and $q=\sum n_i$ or
equivalently
\begin{equation}
\begin{array}{lcr}
m_3=p-(m_1+m_2)\\
n_3=q-(n_1+n_2) .
\end{array}
\end{equation}
This leads to consider the following transformations
\begin{equation}
\begin{array}{lcr}
\sum\limits_{\stackrel{m_in_i}{i=1,2,3}}&\rightarrow&\sum\limits_{pq}\sum\limits_{\stackrel{m_i,n_i}{i=1,2}}\\
\omega(\{m_i\},\{n_i\})_{i=1,2,3}&\rightarrow& \omega(\{m_i\},\{n_i\},p,q)_{i=1,2}=\widetilde{\omega}\\
\epsilon_{123}&\rightarrow&\epsilon_{12\bar{3}}\equiv\widetilde{\epsilon}\\
F_{(123)}&\rightarrow& F_{(12\bar{3})} = {\widetilde F}_{sym} ,
\end{array}
\end{equation}
 
where $\epsilon_{12\bar{3}}$ is the Levi-Cevita tensor introduced previousely
and which is given by\\
\begin{equation}
\begin{array}{lcr}
\epsilon_{12\bar{3}} = \epsilon_{(m_1n_1)(m_2n_2)(pq)}.
\end{array}
\end{equation}
With these simple transformations, the equations of motion (28) become
\begin{equation}
\begin{array}{lcr}
\partial_{\sigma}^2X_{pq} = \sum\limits_{\stackrel{m_i,n_i}{i=1,2}}
8 \widetilde{\epsilon} \widetilde{\omega}\widetilde F_{sym} X_{pq}(\sigma) ,
\end{array}
\end{equation}
with
\begin{equation}
\begin{array}{lcr}
\widetilde{\omega}=(m_2n_1-m_1n_2)[p(n_1+n_2)-q(m_1+m_2)] .
\end{array}
\end{equation}

Note by the way that, as $\widetilde{\omega}$ is antisymmetric with respect to
the following mapping\\
\begin{equation}
\begin{array}{lcr}
(m_1,n_1)&\rightarrow&(m_2, n_2) ,
\end{array}
\end{equation}
${\epsilon_{12\bar{3}}}\widetilde{\omega}$ remains invariant with respect to
the above change of integers values $(m_i, n_i), i=1,2$.
We have\\
\begin{equation}
\begin{array}{lcr}
\epsilon_{12\bar{3}} \omega (\{m_i\},\{n_i\},p,q)
\equiv \epsilon_{21\bar{3}} \omega ({m_1 \rightarrow m_2},{n_1 \rightarrow n_2},p,q)  
\end{array}
\end{equation}
since
\begin{equation}
\begin{array}{lcr}
\epsilon_{12\bar{3}} = -\epsilon_{21\bar{3}}\\
{\omega}(m_i,n_i,p,q)= - {\omega}{({m_1 \rightarrow m_2},
{n_1 \rightarrow n_2},p,q)}
\end{array}
\end{equation}

Next, denoting by $\Omega$ the following function
\begin{equation}
\Omega(p,q)=
\sum\limits_{\stackrel{m_i,n_i}{i=1,2}}8{\widetilde{\epsilon}}
\widetilde{\omega}{\widetilde F}_{sym},
\end{equation}
whose convergence is closely connected to how we can choice the function
${\widetilde F}_{sym}$.\\
In fact the introduction of this symmetric function in our ansatz is really
a crucial step as we can easily check that for arbitrary values of $p, q$ the
following function
\begin{equation}
\begin{array}{lcr}
{\Omega}'(p,q) = \sum\limits_{\stackrel{m_i,n_i}{i=1,2}}8
{\widetilde{\epsilon}} \widetilde{\omega},
\end{array}
\end{equation}
always diverge. To avoid this divergency problem one should then consider
${\widetilde F}_{sym}$ to play the role of the regularisation function which
justify in some sense the introduction of this function in our ansatz (48).
Following this discussion, a natural choice of this regularisation function
is given by
\begin{equation}
\begin{array}{lcr}
\widetilde F_{sym} = F_{(12\bar{3})}= exp(-\vert{f_{12\bar{3}}}\vert),
\end{array}
\end{equation}
where $f_{12\bar{3}}$=$f(\{m_i\},\{n_i\},p,q)$ is some function of the integer
values $m_i, n_i, p, q$ required to be either symmetric or antisymmetric with respect
to permutations of indices $1 \equiv (m_1 n_1)$, $ 2 \equiv (m_2 n_2)$ and
${\bar 3} \equiv (pq)$.\\
The previous choice of $\widetilde F$ as given in (65) is motivated by our
requirement that the function $\Omega (p,q)$ (63) should converge in the infinte limit of the integer
values $m_i, n_i , i=1,2$.\\

Several realisations of the function $f_{(12\bar{3})}$ can occur. We
give herebellow a typical example namely:

a) The symmetric choice

\begin{equation}
\begin{array}{lcr}
f_{(12\bar{3})} = {\sum \limits_{i=1}^{2}(m_i + n_i)}+p+q.
\end{array}, 
\end{equation}

b) The antiymmetric choice

\begin{equation}
\begin{array}{lcr}
f_{(12\bar{3})} = \epsilon_{12\bar{3}}({\sum \limits_{i=1}^{2}(m_i + n_i)}+p+q).
\end{array}
\end{equation}

The equation of motion (58) reads
\begin{equation}
\begin{array}{lcr}
\partial_{\sigma}^2X_{pq}-
\Omega X_{pq}=0 ,
\end{array}
\end{equation}
whose solution is shown to be
\begin{equation}
\begin{array}{lcr}
X_{pq}(\sigma) = A_{pq}e^{\sqrt{\Omega} \sigma}
+ B_{pq}e^{-\sqrt{\Omega} \sigma} ,
\end{array}
\end{equation}
 where
$A_{pq}, B_{pq} $ are arbitrary constants which should satisfy the following
relations
\begin{equation}
\begin{array}{lcr}
A_{pq} A_{rs}=-A_{rs}A_{pq}\\
B_{pq}B_{rs}=-B_{rs}B_{pq} \\
A_{pq}B_{rs}+B_{rs}A_{pq}=-(A_{rs}B_{pq}+B_{pq}A_{rs}) ,
\end{array}
\end{equation}
originated from the non commutativity properties of $X_{pq}$ (43).
Finally the matrix model membrane solution read as
\begin{equation}
\begin{array}{lcr}
X^{\mu}(\alpha,\beta,\sigma)= \sum \limits_{pq} \gamma^{\mu}
(A_{pq}e^{\sqrt{\Omega} \sigma}
+ B_{pq} e^{-\sqrt{\Omega} \sigma})L_{pq} .
\end{array} 
\end{equation}
The following significant question is in order:
how one can compare or connect our solution with those already found
in the same context in [15-16]?\\
One way to do this, is to discuss at the level of the derived solution (71)
some particular cases related to the constants $A_{pq}$ , $B_{pq}$ and the
regularised number $\Omega (p,q)$. \\

$\bf  1. \Omega > 0 $\\
a)$A_{pq}=B_{pq}$\\
\begin{equation}
\begin{array}{lcr}
X^{\mu}(\alpha,\beta,\sigma)= \sum \limits_{pq} 2\gamma^{\mu} A_{pq}
ch( {\sqrt{\Omega} \sigma})
e^{i(p\alpha+q\beta)} .
\end{array}
\end{equation}
b) $A_{pq} = -B_{pq}$\\
\begin{equation}
\begin{array}{lcr}
X^{\mu}(\alpha,\beta,\sigma)= \sum \limits_{pq} 2\gamma^{\mu}
A_{pq} sh({\sqrt{\Omega} \sigma})e^{i(p\alpha+q\beta)} .
\end{array}
\end{equation}

$\bf 2. \Omega < 0 $  \\
a) $A_{pq} = B_{pq}$\\
\begin{equation}
\begin{array}{lcr}
X^{\mu}(\alpha,\beta,\sigma)= \sum \limits_{pq} 2\gamma^{\mu}
A_{pq}\cos({\sqrt{\Omega} \sigma})e^{i(p\alpha+q\beta)} .
\end{array}
\end{equation}
b) $A_{pq} = -B_{pq}$\\
\begin{equation}
\begin{array}{lcr}
X^{\mu}(\alpha,\beta,\sigma)= \sum \limits_{pq} 2i\gamma^{\mu}
A_{pq}\sin({\sqrt{\Omega} \sigma})e^{i(p\alpha+q\beta)} .
\end{array}
\end{equation}

The above examples correspond then to typical membrane solutions
whose oscillation behaviour is the same for all the values of the
space time index $\mu$, with  $\mu=1,...,9$ once the signe of $\Omega$
as  well as the values of $A_{pq}$ and $B_{pq}$ are fixed.\\  
  
Note also that we can consider from our solution (71), other examples for
which  the oscillation behaviour change when changing the index $\mu$ as
done by the author of [15].  We can then conclude that the solution we
have derived is important in the sense that it exhibits among others
an oscillation behaviour having a striking resemblance with the solution
presented for example in [15]. The claim is to remark that the
coordinates p,q and t used for example in the Kim's work, coincide
respectively with $\alpha,\beta$ and $\sigma$ in our construction. \\

\section{Conclusion}
This paper focus to solve some non-linear differential equations decribing
the membrane in 10 dimensions by means of Poisson bracket formalism.\\

Among the results obtained, we derive the equations of motion associated
with $SU(\infty)$ Yang Mills theory describing the membrane in 8 transverse
directions and show how the second order equations of motion are related to
the first order ones (the Nahm's equations). This provides a way to
emphasise the similarity to the phase space formulation of quantum mechanics
as signalled by the author of [4].\\

We rewrite the Nahm's equations in a simple form reducing then their number
from 14 to 5. This is convenient in the sense that their derivative with
respect to $\sigma$ gives in a natural way the equations of motion.\\

We interpret the Nahm's equations as the vanishing property of some conserved
currents associated with the $ SU(\infty)$ Yang Mills theory and the
correspondence with the equations of motion as the conservation property of
these currents. The parameter $\tau$ is shown to play an important role
in this sense.\\

To solve our problem, which means find an explicit expession for the fields
$X^{\mu}$, we develop an algebraic analysis and propose an ansatz leading
to construct the membrane solution given in (71).\\

In the spirit to compare our solution to some of the well known ones,
one performe special choices on the constants $A_{pq}, B_{pq}$  as well
as  on the regularised number $\Omega (p,q)$ (63)  a fact which leads to
recover a general oscillation behaviour (see our previous examples)
shared by our solution and the other ones already established in
literature. \\

Moreover, we guess that other non trivial solutions can be obtained if
one  forget about the ansatz (48) and know how to solve in general the
non-linear  differential equation (28).

\vspace*{3cm}  {\bf Acknowledgements} \\

Two of the authors M.Hssaini and M.B.Sedra would like to thank the
Abdus Salam International Centre for Theoretical Physics and the
considerable help of the High Energy Section where this work was
accomplished. The authors are especially grateful to Prof. K.S. Narain
for reading the manuscript and for his important remarks. We thank also
M. Alif Postdoc at the mathematics section at ICTP for  useful
conversations. Finally we thank  the PARS program: Phys 27.372/98 CNR,
for scientific help.

\newpage

{\bf References}
\begin{enumerate}
\item[[1]] T.Banks, W.Fischler, S.Shenker, and L.susskind, Pys.Rev D55(1997),5112
\item[[2]] D.Dijkgraaf, E.Verlinde and H.Verlinde, Nucl. Phys. B.500(1997)43-61
\item[[3]] see for instance\\
   T.Banks, Nucl. PhysB.Proc.Supp.62(1998)341-347, 68(1998)261-267 \\
   D.Bigatti and L.susskind,hep-th/9712072\\
   D.Brace and B.Morariu, JHEP02,004(1999)hep-th/9810185\\
   D.Brace, B.Morariu and B.Zumino, Nucl.Phys.B 545(1999)192, hep-th/9810099\\
   I. Benkaddour, M. Bennai, E.Y.Diaf and E.H.Saidi, Class.Quant.Grav. 17(2000)1765
\item[[4]] D.Fairlie, Mod.Phys.lettA13(1998)263
\item[[5]] A.Connes, Academic press(1994)
\item[[6]] A.Connes and M.Rieffel, pp.237 contemp.Math.oper.alg.Math.Phys.62,AMS1987.
\item[[7]] N.Seiberg and E.Witten, JHEP09(1999)032.
\item[[8]] A.Connes,M.R.Douglas and A.Schwarz, JHEP,9802:003\\(1998), hep-th/9711162.
\item[[9]] M.R.Douglas and C.Hull, JHEP9802:008,1998, hep-th/9711165.\\
  Y.K.E.Cheung and M.Krogh, Nucl.Phys.B528(1998)185.\\
  B.Morariu and B.Zumino, hep-th/9807198.
\item[[10]] C.S.Chu and P.M.Ho, Nucl.Phys.B550(1999)151,hep-th/9812219.\\
    N.Nekrasov and A.Schwarz, Commun.Mth.Phys.198(1998)\\689,hep-th/9802068.\\
    F.Lizzi and R.J.Szabo, Chaos Solitons Fractals 10(1999)445-458, hep-th/9712206.
\item[[11]] D.B.Fairlie, hep-th/9806198.
\item[[12]] L.Frappat, E. Ragoucy, P. Sorba, F. Thuillier and H. Hogasen, Nucl.Phys.B334,250 (1991)
\item[[13]] I.Antoniadis,P.Ditsas,E.Floratos and J.Iliopoulos, Nucl.Phys.B300, 549(1988).
\item[[14]] E.H.Saidi, M.B.Sedra and A.Serhani,Phys.Lett.B353(1995)209.\\
     E.H.Saidi, M.B.Sedra and A.Serhani,Mod.Phys.Lett.A vol.10N32(1995)2455.
\item[[15]] N. Kim, Phys. Rev.D 59 (1999) 067901 and hep-th/9808166.\\
\item[[16]] J. Hoppe, hep-th/9702169.
\end{enumerate}
\end{document}